\documentclass[12pt]{article}
\usepackage{amsmath}
\usepackage{graphicx}
\usepackage{enumerate}
\usepackage{natbib}
\usepackage{url} 
\usepackage{authblk}

\newcommand{\blind}{1}

\usepackage{geometry}
\usepackage{amsmath}
\allowdisplaybreaks[4]
\usepackage{float}
\usepackage{mathrsfs}
\usepackage{amsfonts}   
\usepackage{xcolor}
\usepackage{xr}
\usepackage{times}
\usepackage{bm}
\usepackage{colonequals}
\usepackage[plain,noend]{algorithm2e}
\usepackage{graphicx}
\usepackage{caption}
\usepackage{enumerate}
\usepackage{cmll}
\usepackage{geometry}
\usepackage{amsthm}
\usepackage{subcaption}
\usepackage{threeparttable}
\usepackage{booktabs}
\usepackage{multirow}
\usepackage{makecell}
\usepackage{tikz}
\usetikzlibrary{positioning,arrows.meta,quotes}
\usetikzlibrary{shapes,snakes}
\usetikzlibrary{bayesnet}
\tikzset{>=latex}
\tikzstyle{plate caption} = [caption, node distance=0, inner sep=0pt,
below left=3pt and 0pt of #1.south]
\usepackage{hyperref}
\usepackage{enumitem}
\usepackage{titlesec}
\usepackage[nodisplayskipstretch]{setspace}

\makeatletter
\def\UrlAlphabet{%
	\do\a\do\b\do\c\do\d\do\e\do\f\do\g\do\h\do\i\do\j%
	\do\k\do\l\do\m\do\n\do\o\do\p\do\q\do\r\do\s\do\t%
	\do\u\do\v\do\w\do\x\do\y\do\z\do\A\do\B\do\C\do\D%
	\do\E\do\F\do\G\do\H\do\I\do\J\do\K\do\L\do\M\do\N%
	\do\O\do\P\do\Q\do\R\do\S\do\T\do\U\do\V\do\W\do\X%
	\do\Y\do\Z}
\def\UrlDigits{\do\1\do\2\do\3\do\4\do\5\do\6\do\7\do\8\do\9\do\0}
\g@addto@macro{\UrlBreaks}{\UrlOrds}
\g@addto@macro{\UrlBreaks}{\UrlAlphabet}
\g@addto@macro{\UrlBreaks}{\UrlDigits}

\titleformat{\section}
{\normalfont\fontsize{12.5}{12.5}\bfseries}{\thesection}{1em}{}
\titleformat{\subsection}
{\normalfont\fontsize{12}{12}\bfseries}{\thesubsection}{1em}{}

\titlespacing{\section}{0pt}{3pt}{3pt}
\titlespacing{\subsection}{0pt}{1pt}{1pt}
\titlespacing{\subsubsection}{0pt}{1pt}{1pt}

\hypersetup{hidelinks}
\graphicspath{{./figures/}}
\makeatletter

\newtheorem{theorem}{\bf{Theorem}}

\newtheorem{lemma}{\bf{Lemma}}

\newtheorem{condition}{\bf{Condition}}

\newcommand{\var}{{\rm var}}

\newcommand{\bmu}{\mu}

\newcommand{\bh}{h}

\newcommand{\bbOmega}{\Omega}

\newcommand{\diag}{\mathsf{diag}}

\newcommand{\tG}{\widetilde{G}}
\newcommand{\tU}{\widetilde{U}}
\newcommand{\tOmega}{\widetilde{\Omega}}
\newcommand{\tM}{\widetilde{M}}

\def\T{{\mathrm{\scriptscriptstyle T} }}

\begin{document}
	\pagenumbering{gobble}

	\def\spacingset#1{\renewcommand{\baselinestretch}%
		{#1}\small\normalsize} \spacingset{1}

	
	\if1\blind
	{
		\title{\bf \fontsize{13}{13}\selectfont Moment-assisted subsampling method for Cox proportional hazards model with large-scale data
		}
	\author{\normalsize  Miaomiao Su and Ruoyu Wang
	\thanks{Miaomiao Su is a lecturer in the School of Science at Beijing University of Posts and Telecommunications ({\em smm@bupt.edu.cn}). Ruoyu Wang is a postdoctoral fellow in the Department of Biostatistics at Harvard T.H. Chan School of Public Health ({\em ruoyuwang@hsph.harvard.edu}).   This work was supported by fundamental research funds from the Beijing University of Posts and Telecommunications (No.2023RC47) and the Key Laboratory of Mathematics and Information Networks (Beijing University of Posts and Telecommunications), Ministry of Education, China.}}
\date{}
\maketitle
} \fi

\if0\blind
{
\bigskip
\bigskip
\bigskip
\begin{center}
	{\bf \fontsize{13}{13}\selectfont Moment-assisted subsampling method for Cox proportional hazards model with large-scale data}
\end{center}
\medskip
} \fi

\bigskip
\begin{abstract}
The Cox proportional hazards model is widely used in survival analysis to model time-to-event data.
However, it faces significant computational challenges in the era of large-scale data, particularly when dealing with time-dependent covariates. 
This paper proposes a moment-assisted subsampling method that is both statistically and computationally efficient for inference under the Cox model. This efficiency is achieved by integrating the computationally efficient uniform subsampling estimator and whole data sample moments that are easy to compute even for large datasets. 
The resulting estimator is asymptotically normal with a smaller variance than the uniform subsampling estimator.
Additionally, we derive the optimal sample moment for the Cox model that minimizes the asymptotic variance in Loewner order. 
With the optimal moment, the proposed estimator can achieve the same estimation efficiency as the whole data-based partial likelihood estimator while maintaining the computational advantages of subsampling. 
Simulation studies and real data analyses demonstrate the promising finite sample performance of the proposed estimator in terms of both estimation and computational efficiency.
\end{abstract}

\noindent%
\vfill

\newpage
\pagenumbering{arabic}
\spacingset{1.7} 
	\section{Introduction}

The semiparametric Cox proportional hazards model \citep{Cox1972} is widely used in survival analysis to study time-to-event data, such as biological death, mechanical failure, or credit default. 
A commonly used method for estimating the Cox regression parameter is the partial likelihood maximization \citep{cox1975partial}. However, when covariates are time-dependent, the computational complexity of this method increases quadratically with the sample size, making it time-consuming for large datasets (see Table \ref{tab: computational complexity}, assuming that sorting $n$ numbers has a time complexity $O(n\log n)$, where $n$ is the whole data size). While this is not a significant issue in classical statistics, it becomes computationally burdensome when applied to large datasets.

To mitigate the challenges posed by storage and computational demands, one solution is to adopt a divide-and-conquer (DAC) strategy. This approach leverages parallel computing by partitioning the whole dataset into several subsets, performing analyses on each subset independently, and then aggregating the results to obtain a final estimator \citep{Mcdonald2009dist,Zhang2013dist,Lee2017dist,battey2018distributed,tang2020distributed}. In recent years, several researchers have extended the DAC strategy to survival analysis.
For instance, \cite{wang2021fast} proposed a DAC algorithm based on linearizations for the Cox proportional hazards model. 
\cite{wang2022multivariate} introduced a weighted DAC method that computes the partial likelihood estimator on each subset and then combines these estimates using a weighted average. The existing DAC approaches are computationally efficient on distributed computing platforms. However, the high costs and limited accessibility of these platforms necessitate the development of methods that are feasible on common personal computers.

Subsampling, a technique that conducts statistical inference on a subset of the whole data, is becoming increasingly popular in recent years. This approach significantly reduces the computational burden brought by large datasets, requires few computing resources, and allows for data analysis on daily equipment, such as laptops, thereby greatly facilitating research. Subsampling is particularly suitable for data preprocessing and exploratory data analysis during the initial stages of research, where researchers often engage in numerous trials to understand the data and develop models. Subsampling methods are especially valuable in the computationally intensive tasks of comparing and debugging different methods during model building.

Uniform subsampling, the simplest form of subsampling, randomly draws a subset from the whole data and estimates based on the subset. While this method is computationally efficient, it often results in relatively low estimation efficiency. To alleviate this, many existing subsampling methods design non-uniform subsampling probabilities (NSP), prioritizing subsamples that are  informative about the parameters of interest \citep{Drines2006Subsampling, Fithian2014Subsampling, Wang2018Subsampling, Yu2022Subsampling, Wang2020LikeliEfficient, Ai2021Subsampling}. The asymptotically optimal NSP-based subsampling (Aopt) method, designed to minimize the trace of the asymptotic variance of the resulting estimator, is particularly popular and has been extensively studied. Extensions of this method have been explored for the additive model by \cite{zuo2021opt} and for the Cox proportional hazards model by \cite{zhang2024optimal}. Additionally, \cite{keret2023rare} developed a modified Aopt method for Cox regression with rare events, incorporating all observed failures.	
Despite its improvements in estimation efficiency, the Aopt method does not guarantee uniform improvement across all parameter components. Moreover, computing the optimal subsampling probabilities can be time-consuming, especially when covariates are time-dependent.

Recently, \cite{wang2024oses} proposed a one-step efficient score (OSES) method under the Cox model. The OSES estimator can achieve the same estimation efficiency as the full-data estimator; however, its computational complexity can be more demanding compared to the Aopt method when covariates are time-dependent due to the need to calculate score function values over the entire dataset  (See Table \ref{tab: computational complexity} and note that the pilot subsample size $r_{0}$ used for the calculation of the optimal NSP can be much smaller than the subsample size $r$.)
In this paper, we assume that all iterative algorithms are linearly convergent for the convenience of discussing computational complexity. 
The computational complexity of the OSES estimator grows polynomially as the subsample size $r$ increases when the covariate is time-dependent.
Moreover, the asymptotic results of \cite{wang2024oses} require the subsample size $r$ to be much larger than $n^{1/2}$. The computing time can be substantial when a large subsample size is taken to meet this condition. On the other hand, numerical experiments demonstrate that the finite sample performance of the OSES estimator is unstable when the subsample size is small.

\begin{table}[htbp]	
	\caption{\label{tab: computational complexity}Computational complexity of different estimation methods. $n$: whole data size, $r$:  subsample size, $r_{0}$: pilot subsample size.}
	\centering
	\begin{tabular}{ccccccccccccccc}
		\toprule
		Method&Covariate&Computational Complexity\\
		\midrule
		Whole &Time-independent  &  $O(n\log n)$ \\
		&Time-dependent  &  $O(n^{2}\log n)$ \\
		\specialrule{0em}{-4pt}{-4pt}\\
		Aopt&Time-independent  &    $O(n\log r_{0})$ \\
		&Time-dependent &    $O(n\log r_{0} +r_{0}n+r^{2}\log r)$ \\
		\specialrule{0em}{-4pt}{-4pt}\\	 
		OSES&Time-independent  &    $O(n\log n)$ \\
		&Time-dependent  &   $O(n\log n +rn+r^{2}\log r)$ \\
		\specialrule{0em}{-4pt}{-4pt}\\	 
		MCox&Time-independent  &   $O(n +r\log r)$ \\
		&Time-dependent  &   $O(n +r^{2}\log r)$ \\
		\bottomrule
	\end{tabular}
\end{table}

This paper develops a moment-assisted subsampling method for the Cox proportional hazards model, referred to as MCox. The MCox method is computationally more efficient than most existing methods and enjoys asymptotic guarantees without restrictions on the rate at which $r$ goes to infinity. See Table \ref{tab: computational complexity} for a comprehensive comparison of the computational complexity of different methods. The MCox method is motivated by the fact that whole data sample moments, defined as the empirical average of a known moment function vector over the entire dataset, are usually informative for the parameter of interest and easy to compute even for large datasets. The MCox method incorporates whole data-based sample moments using the generalized method of moments and adopts a one-step linear approximation to derive the MCox estimator in explicit form.  Under the condition that $r^{2}\log(r) = O(n)$, the time complexity of the MCox estimator scales linearly with the sample size $n$, significantly reducing the computation burden of the whole data-based partial likelihood estimator.  
For any given moment function, the MCox estimator is statistically more efficient than the uniform subsampling estimator for estimating each component of the parameter of interest.  The efficiency improvement depends on the choice of the incorporated sample moment. We derive the optimal moment function for the Cox model and show that when the optimal moment is used, the MCox estimator can achieve the same estimation efficiency as a whole data-based estimator.

It is worth mentioning that the idea of incorporating whole data sample moments has been employed in \cite{su2024moment} to improve the subsampling estimator under a parametric conditional density model. In the parametric model considered by \cite{su2024moment}, the estimating equation is a simple sum of independent terms. However, in the Cox model, observations may contribute multiple terms to the sum, introducing non-trivial dependencies. Consequently, the method in \cite{su2024moment} cannot be directly applied to the Cox model without addressing these complexities.

The rest of this paper is organized as follows. In Section \ref{sec: method}, we introduce the MCox method, establish the asymptotic properties of the resulting estimator, and discuss the determination of the moment function. Sections \ref{sec: simulation} and \ref{sec: application} present simulation studies and a real data application, respectively, to demonstrate the promising finite-sample performance of the MCox estimator. Theoretical derivations are included in Appendix A.

\section{Methodology}\label{sec: method}
\subsection{Model Setup}

In many biomedical applications, the outcome of interest is the occurrence of death or cancer.
The occurrence time is referred to as the failure time \citep{kalbfleisch2011statistical}, which is frequently subject to incomplete observations due to right-censoring.
Let $T$ denote the failure time and $X$ a $p$-dimensional vector of possibly time-dependent covariates.
We assume that the failure time $T$ follows the Cox proportional hazards model \citep{Cox1972}
\begin{equation*}
	\lambda(t; X) = \lambda_{0}(t)e^{\beta_{0}^{\T}X(t)},
\end{equation*}
where $\beta_{0}$ is an unknown regression parameter of interest and $\lambda_{0}(t)$ is an unspecified baseline hazard function.
In practice, the failure time is possibly right-censored. We use $C$ to denote the censoring time, $Y=\min\{T,C\}$ the observed time, and $\Delta=I(T<C)$ the failure indicator. Assume throughout the paper that the censoring time $C$ is independent of the failure time $T$ conditional on the covariate $X$.
Let $\{(Y_{i},X_{i},\Delta_{i})\}_{i=1}^{n}$ be $n$ independently and identically distributed observations of $(Y,X,\Delta)$, where $X_{i}$ is a possibly time-dependent covariate observed on $[0,Y_{i}]$.
Define $N(t) = I(Y\leq t,\Delta=1)$.
Based on the observed data, the partial likelihood estimator $\hat{\beta}$ for $\beta_{0}$ can be obtained by maximizing the following  log-partial likelihood function \citep{cox1975partial}
\begin{equation*}
	\hat{l}(\beta) 
	= \frac{1}{n}\sum_{i=1}^{n}\int_{0}^{\infty}\left[\beta^{\T}X_{i}(t)-\log\left\{\sum_{j=1}^{n}I(Y_{j}\geq t)e^{\beta^{\T}X_{j}(t)}\right\}\right]dN_{i}(t).
\end{equation*}
Let  $a^{\otimes0}=1$, $a^{\otimes1}=a$, and $a^{\otimes2}=aa^{\T}$ for a column vector $a$.
Then, let $\hat{S}^{(l)}(t,\beta)=n^{-1}\sum_{j=1}^{n}I(Y_{j}\geq t)e^{\beta^{\T}X_{j}(t)}{X_{j}(t)}^{\otimes l}$ for $l=0,1,2$ and $\hat{X}(t,\beta) = \hat{S}^{(1)}(t,\beta)/\hat{S}^{(0)}(t,\beta)$.
Then, the score function is 
\begin{equation*}
	\begin{split}
		\widehat{U}(\beta) 
		& = \frac{1}{n}\sum_{i=1}^{n}\int_{0}^{\infty}\left\{X_{i}(t)-\hat{X}(t,\beta)\right\}dN_{i}(t)\\
		& = \frac{1}{n}\sum_{i=1}^{n}\int_{0}^{\infty}\left\{X_{i}(t)-\hat{X}(t,\beta)\right\}dM_{i}(t,\beta),
	\end{split}
\end{equation*}
where $M_{i}(t,\beta) = N_{i}(t)-\int_{0}^{t}I(Y_{i}\geq u)\exp\{\beta^{\T}X_{i}(u)\}\lambda_{0}(u)du$. The information matrix is given by
\begin{equation*}
	\widehat{\Sigma}(\beta) =  \frac{1}{n}\sum_{i=1}^{n}\int_{0}^{\infty}\left\{\hat{S}^{(2)}(t,\beta)/\hat{S}^{(0)}(t,\beta)-\hat{X}(t,\beta)^{\otimes2}\right\}dN_{i}(t).
\end{equation*}

The	Newton-Raphson method is the routine algorithm for solving this problem, and it is also the default optimizer used in the coxph function of the R package  \texttt{Survival} \citep{r2013r,therneau2015package} and the lifelines package in Python  \citep{davidson2019lifelines}.
The computational time of the Newton-Raphson iterative algorithm for obtaining $\hat{\beta}$ depends on the calculation of $\widehat{U}(\beta)$ and $\widehat{\Sigma}(\beta)$ \citep{wang2024oses}. When the covariates are time-independent, the time complexity of calculating $\widehat{U}(\beta)$ and $\widehat{\Sigma}(\beta)$  in each iteration is $O(n\log n)$. However, when the covariates are time-dependent, the complexity increases to $O(n\log n + n^{2})$. 
The optimization procedure usually requires $O(\log n)$ iterations to achieve the desired accuracy.
In this case, as the sample size $n$ increases, the computation of the whole data-based partial likelihood estimate $\hat{\beta}$ becomes time-consuming. 
In this paper, we develop the moment-assisted subsampling method for the Cox proportional hazards model to reduce the computational burden while maintaining high estimation efficiency.

\subsection{Moment-assisted Subsampling Method}

Suppose $\{(Y_{i_{k}},\Delta_{i_{k}},X_{i_{k}})\}_{k=1}^{r}$ is a uniform Poisson subsample drawn from the whole data, where  $r$ is the expected subsample size.
The uniform subsampling estimator $\tilde{\beta}_{\rm uni}$ can be obtained by solving the following subsample-based score estimating equation
\begin{equation}\label{eq:sub}
	\tU(\beta) = \frac{1}{r}\sum_{k=1}^{r}\int_{0}^{\infty}\left\{X_{i_{k}}(t)-\tilde{X}(t,\beta)\right\}dN_{i_{k}}(t)=0,
\end{equation}
where $\tilde{X}(t,\beta) = \tilde{S}^{(1)}(t,\beta)/\tilde{S}^{(0)}(t,\beta)$ and $\tilde{S}^{(l)}(t,\beta)=r^{-1}\sum_{j=1}^{r}I(Y_{i_{j}}\geq t)e^{\beta^{\T}X_{i_{j}}(t)}{X_{i_{j}}(t)}^{\otimes l}$ for $l=0,1,2$.
The information matrix based on the subsample can be given by \begin{equation*}
	\widetilde{\Sigma}(\beta)=  \frac{1}{r}\sum_{k=1}^{r}\int_{0}^{\infty}\left\{\frac{\tilde{S}^{(2)}(t,\beta)}{\tilde{S}^{(0)}(t,\beta)}-\tilde{X}(t,\beta)^{\otimes2}\right\}dN_{i_{k}}(t).
\end{equation*}
The uniform subsampling estimator $\tilde{\beta}_{\rm uni}$ can be quickly computed when the subsample size $r$ is small.
However, it suffers from low estimation efficiency because only a small part of the data is used.     
In this paper, we aim to develop a subsampling method that can reduce the computational of large-scale data while achieving high estimation efficiency.

Suppose $h(\cdot)$ is a function vector of $Z=(Y,\Delta,X)$ and $\mu_{0} = E[h(Z)]$. 
Note that the sample average $\hat{\mu} = n^{-1}\sum_{i=1}^{n}h(Z_{i})$ is usually easy-to-compute even for large datasets.
In addition, $E\{h(Z)-\hat{\mu}\}=0$ implicitly encodes information about $\beta_{0}$. Therefore, we propose to improve the estimation efficiency of $\tilde{\beta}_{\rm uni}$ by utilizing the whole data-based sample moment $\hat{\mu}$.
Based on the subsample, we consider the  easy-to-compute auxiliary estimating function  $r^{-1}\sum_{k=1}^{r}h(Z_{i_{k}})-\hat{\mu}$ and combine it with the estimating function $\tU(\beta)$ in \eqref{eq:sub} to estimate $\beta$.
An efficient way that can achieve this and avoid the over-identified problem is the generalized method of moments \citep{hansen1982}.
Specifically, let 
\begin{equation}
	\tilde{g}(\beta) = \frac{1}{r}\sum_{k=1}^{r}
	\begin{pmatrix}
		\int_{0}^{\infty}\left\{X_{i_{k}}(t)-\tilde{X}(t,\beta)\right\}dN_{i_{k}}(t)\\
		h(Z_{i_{k}})-\hat{\mu}
	\end{pmatrix}.
\end{equation}
Then, we can minimize 
\begin{equation}\label{eq: GMM objective}
	\tilde{g}(\beta)^{\T}\tOmega^{-1} \tilde{g}(\beta)
\end{equation}
to obtain an estimator of $\beta_{0}$, where 
\begin{equation*}
	\tOmega = 
	\begin{pmatrix}
		\tOmega_{11} &\tOmega_{12}\\
		\tOmega_{21} &\tOmega_{22}\\
	\end{pmatrix}
\end{equation*}
is an estimation of the asymptotic variance of $\tilde{g}(\beta_{0})$ with
\begin{equation*}
	\tOmega_{11} = \frac{1}{r}\sum_{k=1}^{r}\left[\int_{0}^{\infty}\left\{X_{i_{k}}(t)-\tilde{X}(t,\tilde{\beta}_{\rm uni})\right\}d\tM_{i_{k}}(t,\tilde{\beta}_{\rm uni})\right]^{\otimes2},
\end{equation*}
\begin{equation*}
	\tOmega_{12} = \tOmega_{21}^{\T} = \frac{1}{r}\sum_{k=1}^{r}(1-r/n)\left[\int_{0}^{\infty}\left\{X_{i_{k}}(t)-\tilde{X}(t,\tilde{\beta}_{\rm uni})\right\}d\tM_{i_{k}}(t,\tilde{\beta}_{\rm uni})\right]\left\{h(Z_{i_{k}})-\tilde{\mu}\right\},
\end{equation*}
\begin{equation*}
	\tOmega_{22}= \frac{1}{r}\sum_{k=1}^{r}(1-r/n)\left(h(Z_{i_{k}})-\tilde{\mu}\right)^{\otimes2},
\end{equation*}
$d\tM_{i}(t,\beta) = dN_{i}(t)-I(Y_{i}\geq t)\exp\{\beta^{\T}X_{i}(t)\}d\widetilde{\Lambda}_{0}(t)$, $$\widetilde{\Lambda}_{0}(t) =\sum_{j=1}^{r}\frac{\Delta_{i_{j}}}{\sum_{l=1}^{r}I(Y_{i_{l}}\geq Y_{i_{j}})\exp(\tilde{\beta}_{\rm uni}^{\T}X_{i_{l}}(Y_{i_{j}}))},$$
and $\tilde{\mu} = r^{-1}\sum_{k=1}^{r}h(Z_{i_{k}})$. 
The minimizor of \eqref{eq: GMM objective} has a closed form if $\tilde{g}(\beta)$ is a linear function of $\beta$ while there is generally no closed form when $\tilde{g}(\beta)$ is nonlinear. To accelerate the computation, we approximate $g(\beta)$ in \eqref{eq: GMM objective} using a linear function $\tilde{g}+\tG(\beta-\tilde{\beta}_{\rm uni})$ and solve the resulting minimization problem to obtain the estimator, where $\tilde{g} = \tilde{g}(\tilde{\beta}_{\rm uni}) = (0^{\T},\tilde{g}_{2}^{\T})^{\T}$, $\tilde{g}_{2} = r^{-1}\sum_{k=1}^{r}\{h(Z_{i_{k}})-\hat{\mu}\}$, and $\tG = (-\widetilde{\Sigma}(\tilde{\beta}_{\rm uni})^{\T},0^{\T})^{\T}$.
The resulting moment-assisted subsampling estimator has the closed form
\begin{equation}
	\begin{split}
		\tilde{\beta}_{\rm MCox} 
		&= \tilde{\beta}_{\rm uni} - (\tG^{\T}\tOmega^{-1}\tG)^{-1}\tG^{\T}\tOmega^{-1}\tilde{g}\\
		& = \tilde{\beta}_{\rm uni} - \widetilde{\Sigma}(\tilde{\beta}_{\rm uni})^{-1}\tOmega_{12}\tOmega_{22}^{-1}\tilde{g}_{2}.
	\end{split}
\end{equation}

For any given moment function $h$, the computational complexity of $\hat{\mu}$ is $O(n)$.
When covariates are time-dependent, the computational complexity of the estimator $\tilde{\beta}_{\rm MCox}$ is $O(n+r^{2}\log r)$, which is significantly lower than the computational complexity $O(n^2\log n)$ associated with the whole data-based partial likelihood estimator. 
The computational complexity of $\tilde{\beta}_{\rm MCox}$ is also lower than that of both the optimal NSP-based subsampling estimator \citep{zhang2024optimal} and the OSES estimator \citep{wang2024oses}. Please refer to Table \ref{tab: computational complexity} for a comprehensive comparison.

\subsection{Asymptotic Properties}

Let $s^{(k)}(t,\beta) = E\{I(Y\geq t)e^{\beta^{\T}X(t)}X(t)^{\otimes k}\}$ for $k=0,1,2$, $\bar{X}(t,\beta) = s^{(1)}(t,\beta)/s^{(0)}(t,\beta)$, 
$$\Sigma(\beta) = E\left[\int_{0}^{\infty}\left\{\frac{s^{(2)}(t,\beta)}{s^{(0)}(t,\beta)}-\bar{X}(t,\beta)^{\otimes2}\right\}dN(t)\right],$$
and $\Sigma_{0} = \Sigma(\beta_{0})$.
Define the population score function
$$\psi(Z;\beta) = \int_{0}^{\infty}\left\{X(t)-\frac{s^{(1)}(t,\beta)}{s^{(0)}(t,\beta)}\right\}dM(t,\beta),$$
and the covariances $\Omega_{11}= E\left[\psi(Z;\beta_{0})^{\otimes2}\right]$, $\Omega_{12}=\Omega_{21}^{\T} = E\left[(1-r/n)\psi(Z;\beta_{0})\{h(Z)-\mu_{0}\}\right]$
and
$\Omega_{22} = E\left[(1-r/n)\{h(Z)-\mu_{0}\}^{\otimes2}\right].$
For any matrix $A$, let $\sigma_{\min}(A)$ and $\sigma_{\max}(A)$ be the minimal and maximal singular values of $A$, respectively.
To establish the asymptotic properties of the MCox estimator $\tilde{\beta}_{\rm MCox}$, we invoke the following regularity conditions.

\begin{condition}\label{condi: mple normal consist}
	(i) $X(t)$ has finite total variation over $[0,\tau]$, where $\tau<\infty$ is the end of the study; (ii) $\int_{0}^{\tau}\lambda_{0}(t)dt<\infty$; (iii) $P(Y\geq \tau)>0$; (iv) $\Sigma_{0}$ is positive definite.
\end{condition}
\begin{condition}\label{condi: moment}
	(i) $E\{\|\bh(Z)\|^{2}\}<\infty$; 
	(ii) $E\{\|\psi(Z;\beta_{0})\|^{2}\}<\infty$.
\end{condition}
\begin{condition}\label{condi: eigenvalue}
	There exist positive constants $c$ and $C$ such that (i) $c<\sigma_{\min}(\Omega_{11})\leq\sigma_{\max}(\Omega_{11})<C$, $c<\sigma_{\min}(\bbOmega_{22})\leq\sigma_{\max}(\bbOmega_{22})<C$, $c<\sigma_{\min}(\bbOmega_{12})\leq\sigma_{\max}(\bbOmega_{12})<C$; (ii) $r\sigma_{\min}(\Sigma_{0}-\bbOmega_{12}\bbOmega_{22}^{-1}\bbOmega_{21})\to \infty$ as $r\to\infty$.
\end{condition}
\begin{condition}\label{condi: lindeber}
	There exist a positive constant $C$ such that $E\left(\left[|b^{\T}\zeta|/\{\var(b^{\T}\zeta)\}^{1/2}\right]^{2+c}\right)\leq C$ for any vector $b$, where $\zeta = \delta\psi(Z;\beta_{0})-\{\delta-r/n\}\bbOmega_{12}\bbOmega_{22}^{-1}\{\bh(Z)-\bmu_{0}\}$, $\delta$ is an inclusion indicator for $Z$ and $\delta=1$ if $Z$ is included in the subsample, $\delta=0$ otherwise.
\end{condition}

Condition \ref{condi: mple normal consist} is a standard assumption in censored linear regression, commonly used to ensure the asymptotic properties of the partial likelihood estimator \citep{andersen1982cox}. Condition \ref{condi: moment} specifies a regular moment condition and Condition \ref{condi: eigenvalue} imposes requirements on the eigenvalues of certain matrices. Condition \ref{condi: lindeber} is another moment condition required for applying the Lindeberg-Feller central limit theorem, ensuring the asymptotic normality of the proposed estimator.

\begin{theorem}\label{theo: asynormal}
	Under Conditions \ref{condi: mple normal consist} -- \ref{condi: lindeber}, we have
	\begin{equation}
		V_{h}^{-1/2}(\tilde{\beta}_{\rm MCox}-\beta_{0})\stackrel{d}{\to}N(0,I),
	\end{equation}
	where $V_{h} = r^{-1}\Sigma_{0}^{-1}\left(\Omega_{11}-\Sigma_{0}^{-1}\Omega_{12}\Omega_{22}^{-1}\Omega_{21}\Sigma_{0}^{-1}\right)\Sigma_{0}^{-1}$.
\end{theorem}

From Theorem \ref{theo: asynormal}, it is evident that the asymptotic variance $V_{h}$ of the MCox estimator is smaller than  the variance $V_{0} = r^{-1}\Sigma_{0}^{-1}$ of the uniform subsampling estimator for any moment function $h$. This indicates that the MCox estimator achieves higher estimation efficiency compared to the uniform subsampling estimator. Note that the asymptotic variance $V_{h}$ depends on the moment function $h$. In the following, we explore how to determine $h$ for the implementation of the MCox estimator. 

\begin{theorem}\label{th: opt h}
	The asymptotic variance $V_{h}$ attains the minimum $n^{-1}\Sigma_{0}^{-1}$ if and only if the moment function $h$ satisfies $M[h(Z)-E\{h(Z)\}]=\psi(Z,\beta_{0})$ for some matrix $M$.
\end{theorem}

Theorem \ref{th: opt h} establishes the necessary and sufficient conditions for a moment function $h$ to be optimal. When $h$ is the optimal moment function, the asymptotic variance $V_{h}$ achieves its minimum, which equals to the asymptotic variance $n^{-1}\Sigma_{0}^{-1}$ of the whole data-based partial likelihood estimator $\hat{\beta}$.

Note that $h(Z) = \psi(Z,\beta_{0})$ meets the criteria for an optimal moment function. However, it involves unknown parameters. 
We propose to take a pilot subsample of size $r_{0}$, indexed by $\mathcal{J}_{0}$, to estimate the moment function.
Specifically, for practical implementation of $\hat{\mu}$, we utilize the estimated optimal moment function
\begin{equation*}
	\begin{split}
		\tilde{h}^{\rm opt}(Z) &= \tilde{\psi}(Z,\tilde{\beta}_{\rm uni}) \\
		& = \Delta\left\{X(Y)-\frac{\check{S}^{(1)}(Y,\tilde{\beta}_{\rm uni})}{\check{S}^{(0)}(Y,\tilde{\beta}_{\rm uni})}\right\}\\
		&\quad-\sum_{j\in\mathcal{J}_{0}}\left\{X(Y_{j})-\frac{\check{S}^{(1)}(Y_{j},\tilde{\beta}_{\rm uni})}{\check{S}^{(0)}(Y_{j},\tilde{\beta}_{\rm uni})}\right\}I(Y_{j}\leq Y)\exp\{\tilde{\beta}_{\rm uni}^{\T}X(Y_{j})\}d\check{\Lambda}_{0}(Y_{j}),
	\end{split}
\end{equation*}
where $\check{S}^{(l)}(t,\beta)=r_{0}^{-1}\sum_{j\in\mathcal{J}_{0}}I(Y_{j}\geq t)e^{\beta^{\T}X_{j}(t)}{X_{j}(t)}^{\otimes l}$ for $l=0,1,2$, and
$$d\check{\Lambda}_{0}(Y_{j}) =\frac{\Delta_{j}}{\sum_{l\in \mathcal{J}_{0}}I(Y_{l}\geq Y_{j})\exp(\tilde{\beta}_{\rm uni}^{\T}X_{l}(Y_{j}))}.$$
When covariates are time-independent, the time complexity of $\tilde{h}^{\rm opt}$-based $\hat{\mu}$ is $O(n\log r_{0})$. We recommend directly incorporating the estimated optimal $h$ into the MCox estimator.
The computation of $\tilde{h}^{\rm opt}$ is more complex when covariates are time-dependent. One can take the pilot sample size $r_{0}$ to be smaller than the subsample size $r$ to reduce the computational burden for calculating the estimated moment function for the whole data. Our simulation shows that $r_{0} = r^{2/3}\log r$ produces quite promising numerical results. Then, the computational complexity of $\hat{\mu}$ is $O(r_{0}n+n\log r_{0})$, leading to an overall time complexity of $O(r^{2}\log r+r_{0}n+n\log r_{0})$ for the MCox estimator which is lower than that of the OSES estimator. 

Alternatively, to further address computational challenges with time-dependent covariates, one can use a reasonable parametric approximation of the optimal moment function $\psi(Z;\beta_{0})$, such as the score function from the accelerated failure (AFT) model. 
The AFT model is a parametric model introduced by \cite{Cox1972}, primarily used for studying the reliability of industrial products. In the AFT model, covariate effects act multiplicatively on survival time, making it a good alternative to the Cox model when analyzing survival data \citep{wei1992accelerated}. Additionally, if survival times follow a Weibull distribution, the Cox model can be re-parameterized as a Weibull AFT model, and the deceleration factors of the AFT model should correspond to log-transformed hazard ratios \citep{collett2023modelling}. 
This approximation leads to a time complexity of $O(n)$ for calculating $\hat{\mu}$, lower than that based on the estimated optimal moment function $\tilde{h}^{\rm opt}$. Employing the parametric approximation can control the computational complexity of the MCox estimator to $O(r^{2}\log r + n)$.
In addition, the optimal moment function can also be approximated using nonparametric methods such as tree-based approaches \citep{friedman2001greedy} or sieve methods \citep{shen1994convergence} fitted on a subsample. 	
Importantly, regardless of the moment function used, the proposed estimator remains more efficient than the subsampling estimator without incorporating moment information.

\subsection{Connection with the One-step Estimator}\label{subsec: one-step}

We next explore the connections and differences between the proposed estimator and the OSES estimator introduced by \cite{wang2024oses}. For ease of comparison, assume that the OSES estimator is refined using the efficient score calculated on the whole data. In addition, assume the estimated optimal moment function $\tilde{h}^{\rm opt}$ is incorporated in the MCox estimator and the pilot subsample used in $\tilde{h}^{\rm opt}$ is the whole data. 
In this case, some calculations can show that the MCox estimator is $\tilde{\beta}_{\rm MCox} = \tilde{\beta}_{\rm uni} + (1-\alpha)\widetilde{\Sigma}(\tilde{\beta}_{\rm uni})^{-1}\hat{\mu}$, where $\alpha = \hat{\mu}^{\T}(\tOmega_{11}+\hat{\mu}\hat{\mu}^{\T})^{-1}\hat{\mu} \in [0, 1]$. 
This formula resembles that of the OSES estimator $\tilde{\beta}_{\rm oses} = \tilde{\beta}_{\rm uni} + \widetilde{\Sigma}(\tilde{\beta}_{\rm uni})^{-1}\hat{\mu}$.
Note that $\tilde{\beta}_{\rm oses}$ can be obtained by minimizing the function
\begin{equation}\label{eq: os app}
	(\beta-\tilde{\beta}_{\rm uni})^{\T}\widetilde{\Sigma}(\tilde{\beta}_{\rm uni})(\beta-\tilde{\beta}_{\rm uni})/2+\hat{\mu}(\beta-\tilde{\beta}_{\rm uni}),
\end{equation}
which is an approximation of the second-order Taylor expansion of the whole data-based log-partial likelihood function.
The approximation performs well only when $\tilde{\beta}_{\rm uni}$ is close to the whole data-based partial likelihood estimator $\hat{\beta}$.
When the subsample size $r$ is small,  $\tilde{\beta}_{\rm uni}$ may deviate from $\hat{\beta}$, leading to poor finite sample performance of  $\tilde{\beta}_{\rm oses}$. 
This phenomenon is demonstrated in our simulation study. 
In contrast, the MCox estimator uses the adaptive step size $1-\alpha$.
When $\tilde{\beta}_{\rm uni}$ deviates from $\hat{\beta}$,  $\|\hat{\mu}\|$ tends to be large, making $\alpha$ large, which results in a small step size. This is reasonable because \eqref{eq: os app} is not a good approximation and $\widetilde{\Sigma}(\tilde{\beta}_{\rm uni})^{-1}\hat{\mu}$ may not be a good updating step in this case. 
Conversely, if $\tilde{\beta}_{\rm uni}$ is close to $\hat{\beta}$,  $\|\hat{\mu}\|$ tends to be small, resulting in a small $\alpha$. Then, the step size $1 - \alpha$ is close to $1$ and the updates in $\tilde{\beta}_{\rm MCox}$ and $\tilde{\beta}_{\rm oses}$ are similar in this case. 
The above adaptive property enables the proposed method to perform well even when the subsample size is quite small. This is verified by the numerical results.

\section{Simulation Study}\label{sec: simulation}

In this section, we evaluate the finite sample performance of the proposed MCox method through simulations. We consider the estimated optimal moment function $\tilde{h}^{\rm opt}$ and refer to the resulting MCox estimator as MCox-OPT. Given the computational complexity of $\tilde{h}^{\rm opt}$ with time-dependent covariates, we also consider the moment function $\tilde{h}^{\rm app}$ --- the estimated score function of the AFT model. The $\tilde{h}^{\rm app}$-based MCox estimator is denoted as MCox-APP. For comparison, we also calculate the uniform subsampling (UNI) estimator, the Aopt estimator in \cite{zhang2024optimal}, and the OSES estimator in \cite{wang2024oses}. All numerical studies were performed on a Windows server with a 52-core processor and 128GB RAM. The Aopt and OSES are implemented using their respective published codes.

We generate the failure time $T$ from a Cox model with a baseline hazard function $\lambda_{0}(t) = 1$ and true parameter $\beta_{0} = (0.2,0.2,0.1,0.1,0.1)^{\T}$. The censoring time $C$ is generated from a uniform distribution over $(0,c_{0})$ where $c_{0}$ is set to $3.275$ such that the censoring rate is $70\%$. 
Two settings are considered: (1) time-independent covariates $X_{ind}$ generated from a multivariate $t$-distribution with degrees of freedom 10, a mean vector of zeros, and a covariance matrix ${(0.5^{|i-j|})}_{i,j=1,\dots,5}$, and (2) time-dependent covariates $X_{dep} = X_{ind} + t\epsilon$, where $\epsilon$ follows a multivariate normal distribution with a mean vector of zeros and a diagonal covariance matrix $\diag(0.4, 0.4, 0.4, 0.4, 0.4)$.

\begin{figure}[htbp]
	\centering
	\includegraphics[scale=0.4]{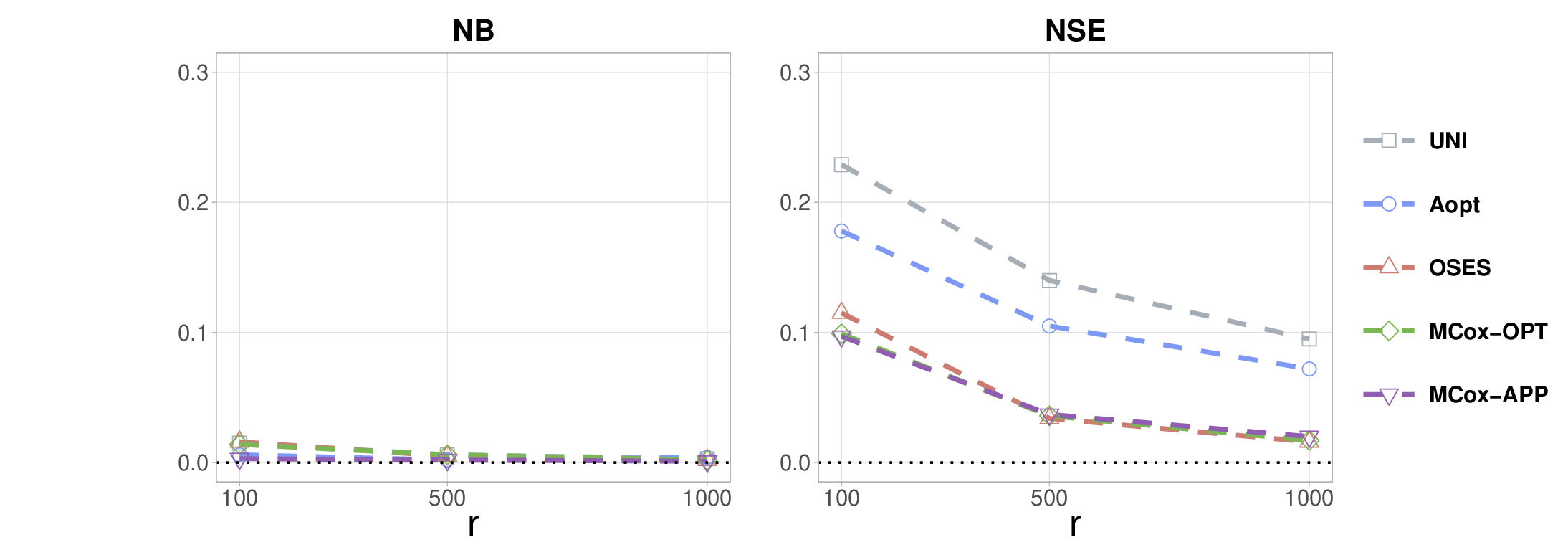}
	\hspace{20mm}
	\caption{The NB and NSE of different subsampling estimators under the Cox model with time-independent covariates and $n=10^{7}$.}
	\label{fig: ti_nb_nse}
\end{figure}

In the first setting (time-independent covariates), the whole data size is $n = 10^7$, and the three subsample sizes $r = 100$, $500$, and $1000$ are considered. 
In addition, we randomly draw a pilot subsample of size $r_{0}=r^{2/3}\log r$ for the implementation of the optimal subsampling probability and the moment functions $\tilde{h}^{\rm opt}$ and $\tilde{h}^{\rm app}$.
Figure \ref{fig: ti_nb_nse} shows the norm of bias (NB) and norm of standard error (NSE) based on 1000 simulations. As $r$ increases, NB and NSE of all estimators decrease, with MCox and OSES estimators outperforming other subsampling estimators in terms of NSE. The MCox and OSES estimators yield significantly lower NSEs than UNI and Aopt estimators across all subsample sizes. The results are consistent with our theoretical results that MCox and OSES estimators can asymptotically achieve the same convergence rate as the whole data-based estimator, which is faster than UNI and Aopt. 

\begin{figure}[htbp]
	\centering
	\includegraphics[scale=0.7]{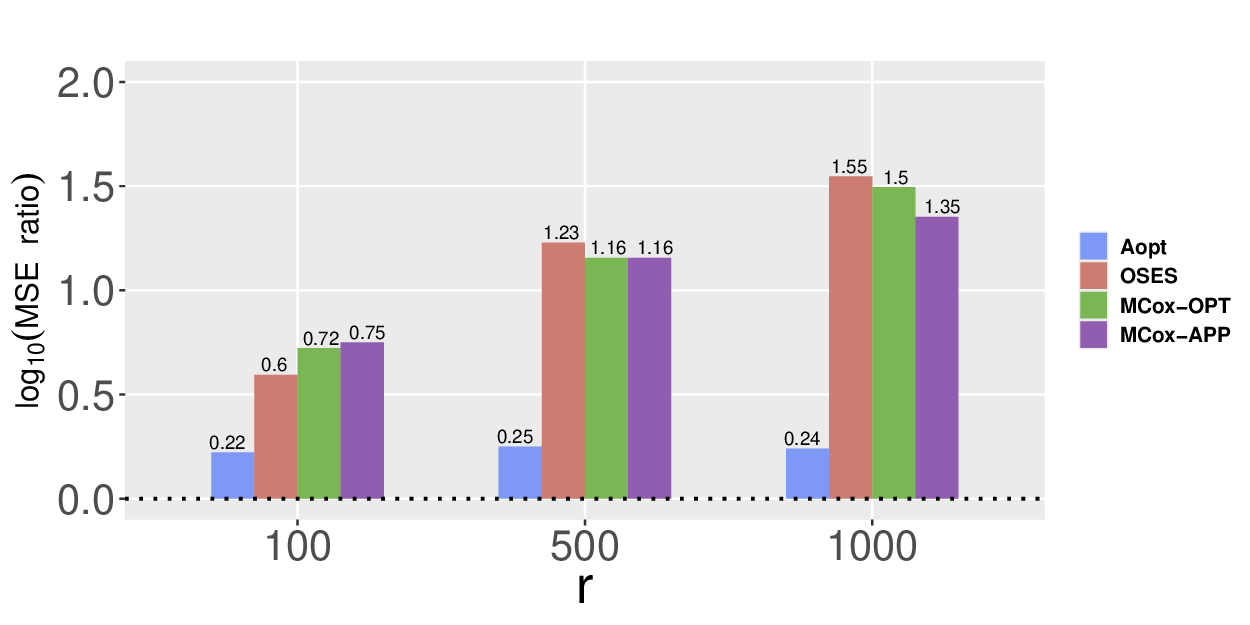}
	\hspace{10mm}
	\caption{The logarithm of the MSE ratio of the UNI estimator to the Aopt, OSES, and MCox estimators under the Cox model, with time-independent covariates and $n=10^{7}$.}
	\label{fig: ti_rela_eff}
\end{figure}

Figure \ref{fig: ti_rela_eff} plots the mean square error (MSE) ratio of UNI estimator to other subsampling estimators, showing that MCox-OPT, MCox-APP, and the OSES estimator have significantly lower MSEs. The results associated with MCox-OPT and MCox-APP confirm the benefits of incorporating whole data-based moment information. 

Table \ref{tab: ti time} presents the computing time of different estimators. For reference, we also include the computing time of the whole-data partial likelihood estimator using the R function \texttt{coxph}. All subsampling estimators take much less computing time than the whole-data estimator, with the UNI estimator being the fastest. However, the UNI estimator suffers from substantial estimation efficiency loss. Although the Aopt, OSES, and MCox estimators require more time than the UNI estimator, they achieve significantly higher estimation efficiency. 
Their computing times show little change as the subsample size increases. This is likely because the whole dataset is much larger than the subsample, so the main computational cost arises from computing the NSP and whole-data sample moments. 
Overall, considering both statistical and computational efficiency, the OSES and MCox estimators are preferable when covariates are time-independent.

\begin{table}[htbp]	
	\caption{\label{tab: ti time}CPU times (in seconds) for different estimators under the Cox model with time-independent covariates and $n=10^{7}$.}
	\centering
	\begin{tabular}{ccccccccccccccc}
		\toprule
		$r$&UNI&Aopt&OSES&MCox-OPT&MCox-APP\\
		\midrule
		$100$ &0.001 &2.803& 5.208 &  4.024 &  1.163\\
		\specialrule{0em}{-4pt}{-4pt}\\
		$500$ &0.003 &2.810& 5.242 &4.218 &1.214\\
		\specialrule{0em}{-4pt}{-4pt}\\	 
		$1000$ &0.006 &2.829& 5.319 &   4.130 &  1.191 \\ 
		\specialrule{0em}{-4pt}{-4pt}\\
		\multicolumn{4}{c}{Whole data-based estimator:  75.14}\\
		\bottomrule
	\end{tabular}
\end{table}

\begin{figure}[htbp]
	\centering
	\includegraphics[scale=0.4]{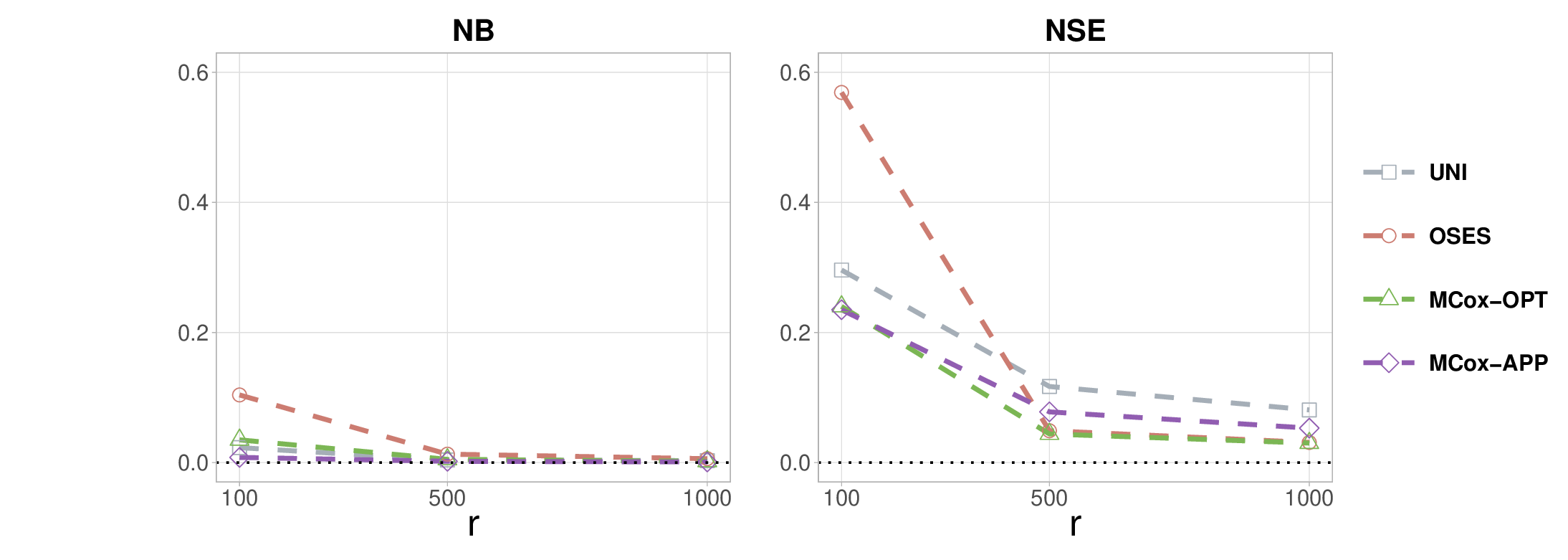}
	\hspace{20mm}
	\caption{The NB and NSE of different subsampling estimators under the Cox model with time-dependent covariate and $n=10^{4}$.}
	\label{fig: tv_nb_nse}
\end{figure}

In the second setting (time-dependent covariates), the computing time for all methods is generally more extensive than in the first setting. To effectively assess both the estimation and computational efficiency, we fix the whole data sample size at $n = 10^{4}$ and vary the subsample size $r$ to be $100$, $500$, and $1000$.
We also randomly draw a pilot subsample of size $r_{0} = r^{2/3}\log r$ for the implementation of the moment functions $\tilde{h}^{\rm opt}$ and $\tilde{h}^{\rm app}$.
The Aopt estimator in \cite{zhang2024optimal} is proposed in the setting where covariates are time-independent and hence the simulation results of Aopt are not presented in the second setting.
Figure \ref{fig: tv_nb_nse} plots the NB and NSE of different estimators based on 1000 simulations. Figure \ref{fig: tv_nb_nse} shows that both NB and NSE decrease as $r$ increases. MCox-APP outperforms the UNI estimator in terms of NSE, which is consistent with the theoretical results in Theorem \ref{theo: asynormal}.
The OSES estimator shows higher NSE than MCox-OPT and MCox-APP when $r = 100$, consistent with the analysis in Section \ref{subsec: one-step}. The good performance of the MCox estimators owes to their adaptive properties, particularly with small subsamples.

\begin{figure}[htbp]
	\centering
	\includegraphics[scale=0.6]{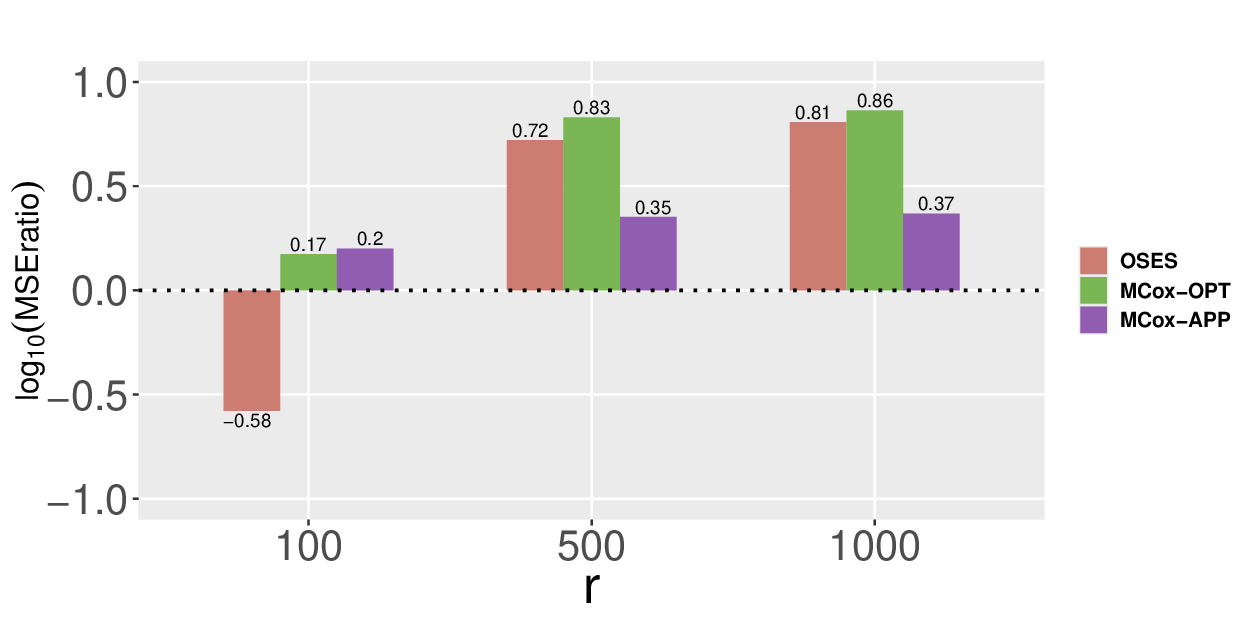}
	\hspace{10mm}
	\caption{The logarithm of the MSE ratio of the UNI estimator to the OSES and MCox estimators under the Cox model with time-dependent covariate and $n=10^{4}$.}
	\label{fig: tv_rela_eff}
\end{figure}

Figure \ref{fig: tv_rela_eff} plots the MSE ratio of the UNI estimator to other subsampling estimators.	From Figure \ref{fig: tv_rela_eff}, it can be seen that when the subsample size is large, MCox-OPT and the OSES estimator have similar performance which is better than that of MCox-APP.
However, when the subsample size is reduced to $r=100$, the MSE of the OSES estimator exceeds that of the UNI estimator.
In contrast, both MCox-OPT and MCox-APP consistently outperform the UNI estimator.

We further present the computing time of different estimators in Table \ref{tab: tv time}.  Table \ref{tab: tv time} shows that all subsampling estimators require significantly less computing time than the whole data-based partial likelihood estimator. Among these, the computational cost of MCox-APP is comparable to that of the UNI estimator. The computing time of MCox-OPT and MCox-APP are both notably shorter than that of the OSES estimator. This computational efficiency makes MCox estimators appealing choices for handling time-dependent covariates, as it offers a good balance between computational efficiency and statistical efficiency.

\begin{table}[htbp]	
	\caption{\label{tab: tv time}CPU times (in seconds) for different estimators under the Cox model with time-dependent covariate and $n=10^{4}$.}
	\centering
	\begin{tabular}{ccccccccccccccc}
		\toprule
		$r$&UNI&OSES&MCox-OPT &MCox-APP\\
		\midrule
		$100$&0.02& 33.39   & 0.22 &   0.02\\
		\specialrule{0em}{-4pt}{-4pt}\\
		$500$&0.25 &37.83    &0.92   & 0.28\\
		\specialrule{0em}{-4pt}{-4pt}\\
		$1000$&1.04 &41.78   & 2.38   & 1.13\\
		\specialrule{0em}{-4pt}{-4pt}\\
		\multicolumn{4}{c}{Whole data-based estimator:  663}\\
		\bottomrule
	\end{tabular}
\end{table}

\section{Real data application}\label{sec: application}

It is estimated that worldwide more than 7.6 million perinatal deaths occur annually, 57\% of which are fetal deaths \citep{conde2000epidemiology}. Increasing maternal age has been identified as a significant risk factor for fetal mortality \citep{Fretts1995}, and the relationship between maternal age and the risk of fetal death has garnered widespread attentions and researches \citep{haavaldsen2010impact,alio2012effect,martin2018births}. Significant changes in socioeconomic, cultural, and policy environments—such as higher education levels, insufficient workplace support, cultural shifts, economic instability, policy restrictions, healthcare challenges, and changes in personal relationship dynamics—have contributed to a sharp increase in childbirth among women aged 35 and older \citep{mills2011people,molina2019delay}.
Understanding how maternal age influences the risk of fetal mortality has profound implications for family structures, labor markets, and the formulation of public policies. 

In this section, we aim to examine whether advanced maternal age increases the risk of fetal death and whether this relationship varies with gestational weeks.
To address these questions, we apply the proposed MCox method to fetal death data from the United States, publicly accessible through the National Center for Health Statistics (NCHS).
The dataset includes 1,930,825 subjects with various demographic details from 1989 to 2022.	
Our analysis focuses on the effects of maternal age on fetal death, with gestational weeks as the failure time. Maternal age is categorized into five groups: 0 (under 20 years), 1 (20–24 years), 2 (25–29 years), 3 (30–34 years), and 4 (35 years and older). It is encoded into four dummy variables $(A_{1},A_{2},A_{3},A_{4})$, with ``under 20 years'' serving as the reference group.
To control for potential confounders, we include four additional covariates: fetal sex ($Z_{1}$, 0: female, 1: male), resident status ($Z_{2}$, 0: resident, 1: non-resident), plurality ($Z_{3}$, 0: single, 1: multiple), and mother's race ($Z_{4}$, 0: white, 1: non-white). Given the small proportion of missing data, we simply drop rows with missing values, resulting in 1,838,675 subjects available for analysis.

To better understand the relationship between maternal age and death risk, we first plot the marginal risk of maternal age on fetal death using a pilot sample, which reveals an approximate quadratic trend (see Figure 1 in the Appendix B).
Motivated by this quadratic pattern, we use Legendre polynomial to model the time-varying coefficients as  $\beta(t) = \beta^{(0)}+2\beta^{(1)}t+\beta^{(2)}(4t^{2}-2)$, where $\beta^{(k)},k=0,1,2$ are 8-dimensional unknown parameters.
Let $\beta^{(k)} = (\beta_{A}^{(k)\T},\beta_{Z}^{(k)\T})^{\T}$, where $\beta_{A}^{(k)}$ and $\beta_{Z}^{(k)}$ correspond to the first four and last four dimensions of $\beta^{(k)}$, respectively. Then the parameters of interest, $\beta_{A}(t) = \beta_{A}^{(0)}+2\beta_{A}^{(1)}t+\beta_{A}^{(2)}(4t^{2}-2)$, represent the effect of maternal age on fetal death as the gestational week varies.
We then consider a Cox proportional hazards model with time-varying coefficients $\beta(t)$.
Let $X=(A_{1},A_{2},A_{3},A_{4},Z_{1},Z_{2},Z_{3},Z_{4})^{\T}$ be the vector of covariates. The failure risk function $\lambda(t;X) =\lambda_{0}(t)e^{X^{\T}\beta(t)} = \lambda_{0}(t)e^{X^{\T}\beta^{(0)}+2X^{\T}\beta^{(1)}t+X^{\T}\beta^{(2)}(4t^{2}-2)}$.

Define $\beta = (\beta^{(0)\T},\beta^{(1)\T},\beta^{(2)\T})^{\T}$ and $X(t) = (X^{\T},2X^{\T}t,X^{\T}(4t^{2}-2))^{\T}$.
This reformulates the Cox proportional hazards model with time-varying coefficients into a Cox proportional hazards model with time-dependent covariates: $\lambda(t;X) = \lambda_{0}(t)e^{\beta^{\T}X(t)}$.
We apply the $\tilde{h}^{\rm app}$- and $\tilde{h}^{\rm opt}$-based MCox methods to analyze the data.
We take the subsample of size $r=6000$ to calculate these estimators. In addition, we randomly draw a pilot subsample of size $r_{0}=r^{2/3}\log r$ for the implementation of the moment functions $\tilde{h}^{\rm opt}$ and $\tilde{h}^{\rm app}$.
The results are summarized in Table \ref{table: fetus}, along with comparisons to results from the uniform subsampling estimator, one-step efficient score estimator and the full data-based partial likelihood estimator.
Metrics in Table \ref{table: fetus} include the average of standard error (ASE) and the symmetric difference (Diff) of the selected covariates in $X(t)$ compared to those identified by the whole data-based partial likelihood estimator at a 0.05 significance level. The symmetric difference quantifies the discrepancies in covariate selection between different estimators and the benchmark (whole data-based estimator).

\begin{table}[htbp]	
	\caption{\label{table: fetus}Comparison of estimation methods for fetal death data}
	\centering
	\begin{tabular}{ccccccccccccc}
		\toprule
		\multicolumn{2}{c}{\text{Whole}} & \multicolumn{2}{c}{\text{UNI}} & \multicolumn{2}{c}{\text{OSES}} & \multicolumn{2}{c}{\text{MCox-APP}}& \multicolumn{2}{c}{\text{MCox-OPT}}\\
		\cmidrule(lr){1-2}\cmidrule(lr){3-4}\cmidrule(lr){5-6} \cmidrule(lr){7-8} \cmidrule(lr){9-10}
		\small{ASE} & \small{Diff} & \small{ASE} & \small{Diff} & \small{ASE} & \small{Diff} & \small{ASE} & \small{Diff}& \small{ASE} & \small{Diff}\\
		\midrule
		0.024 &0 &0.415  &11 &0.024  &0 & 0.323 &6  &0.024 &0\\
		\bottomrule
	\end{tabular}
\end{table}


From Table \ref{table: fetus}, the whole data-based estimator indicates that all covariates --
$X$,$Xt$,and $Xt^{2}$ -- have significant effects on fetal gestational weeks at the 0.05 significance level.  In contrast, the UNI method fails to identify 11 of these covariates.
MCox-APP performs better than the UNI method,  missing only 6 significant covariates. Notably, both MCox-OPT and the OSES method select all significant covariates, exhibiting the same performance as that of the whole data-based estimator.
Additionally, the ASE of the MCox estimators is much smaller than that of the UNI estimator regardless of the moment function used. The ASEs of MCox-OPT and the OSES estimator is comparable to that of the whole data-based partial likelihood estimator.

In terms of computational efficiency, the computing times of the Whole, UNI, OSES, MCox-APP, and MCox-OPT estimators are 896.7, 1.6, 34.6, 2.6, and 14.4 seconds, respectively. While the UNI estimator is the fastest, it suffers from low estimation efficiency. In contrast, MCox-APP is nearly as fast as UNI but offers higher estimation efficiency.
Meanwhile, MCox-OPT achieves the same estimation efficiency as OSES and the whole-data estimators with the shortest computing time. 

\begin{figure}
	\centering
	\includegraphics[scale=0.45]{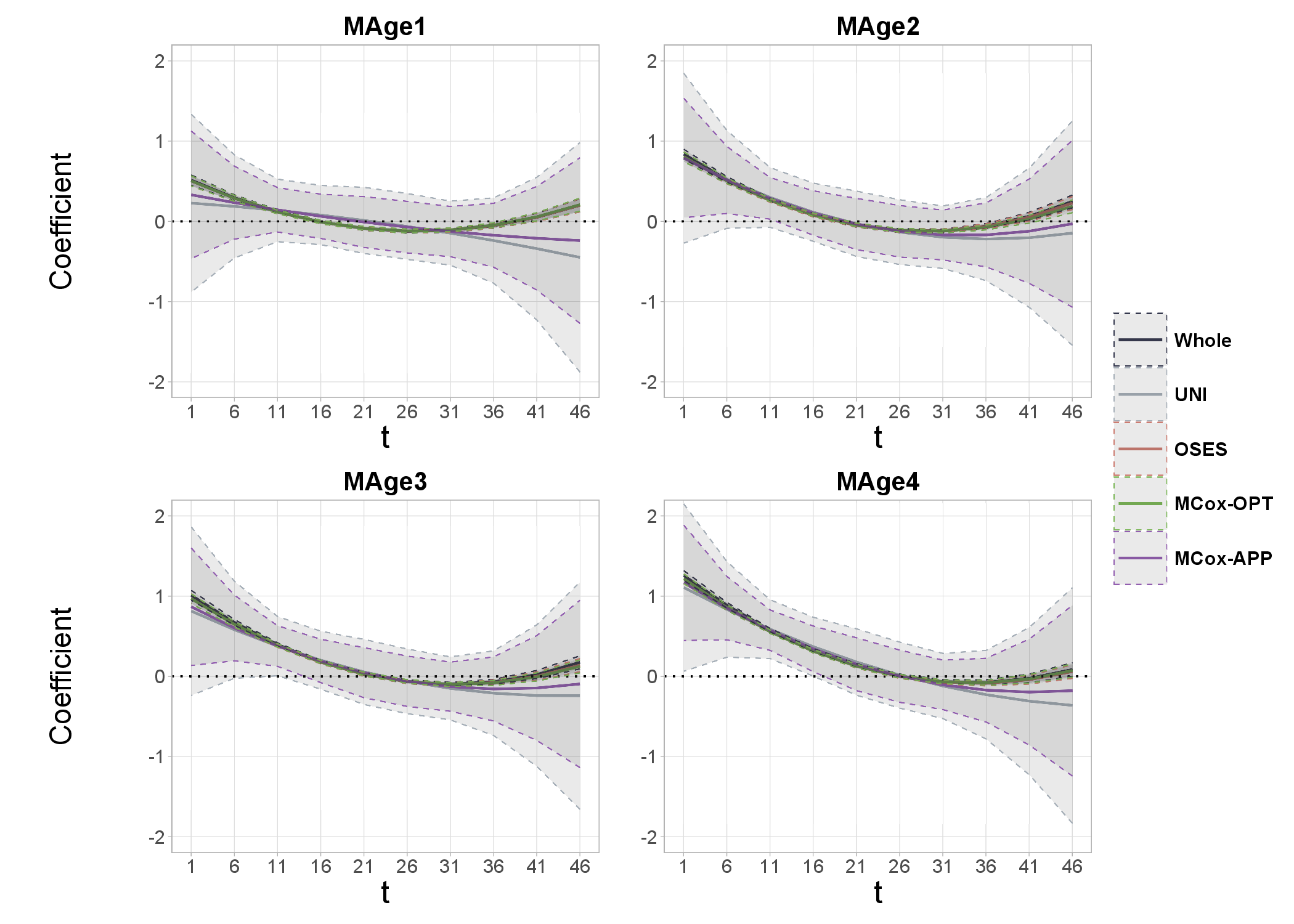}
	\caption{Time-varying coefficients of maternal ages on fetal death with 95\% confidence interval. MAge: maternal age (MAge1: 20-24 years, MAge2: 25-29 years, MAge3: 30-34 years, MAge4: 35 years and older)}
	\label{fig: coef}
\end{figure}

To evaluate the total effect of maternal age, we recover $\beta_{A}(t)$ from the estimates of $\beta_{A }^{(k)},k=0,1,2$ and plot the time-varying coefficients in Figure \ref{fig: coef}. From Figure \ref{fig: coef}, MCox-OPT and the OSES estimator reach the same trend as the whole data-based partial likelihood estimator: the risk of fetal death is significantly higher during early pregnancy compared to later stages for all age groups. Moreover, the risk of fetal death increases with maternal age during early pregnancy.
In late pregnancy, MCox-OPT estimator, the OSES estimator and the whole data-based partial likelihood estimator similarly conclude that the risk of fetal death is higher than during the mid-pregnancy phase across all age groups.  MCox-APP demonstrates better performance than the UNI estimator in terms of both the confidence interval width and alignment with the trend of the whole data-based estimator.

	\section{Concluding Remark}\label{sec: conclusion}

In this paper, we propose the MCox method for Cox regression to address the challenges posed by large-scale data. The method substantially reduces computational burdens, particularly when covariates are time-dependent. By incorporating whole-data sample moments, MCox improves statistical efficiency with minimal additional computation. Beyond its theoretical statistical and computational advantages, the proposed estimator demonstrates strong finite-sample performance, thanks to its initial-estimator-adaptive design discussed in Section \ref{subsec: one-step}. These features make the MCox method a desirable and reliable tool for large-scale data analysis in survival studies.

\newpage

\appendix

\renewcommand{\thesection}{Appendix \Alph{section}}

\section{Proofs of Theorems 1 and 2}
We first establish the following two lemmas that are used in the proofs of  Theorems 1 and 2.
\begin{lemma}\label{lem: consist Sk}
	Under Condition 1, we have
	\begin{equation*}
		\sup_{t\in[0,\tau],\beta}\|\tilde{S}^{(l)}(t,\beta)-s^{(l)}(t,\beta)\|\stackrel{p}{\to}0.
	\end{equation*}
	\begin{proof}
		Under Condition 1, Lemma \ref{lem: consist Sk} can be proved following the same arguments as those used in the proof of Lemma 1 in \cite{wang2024oses}.
	\end{proof}
\end{lemma}

\begin{lemma}\label{lem: remain rate}
	Under Condition 1, we have
	\begin{equation*}
		\frac{1}{r}\sum_{k=1}^{r}\int_{0}^{\infty}\left\{\frac{\tilde{S}^{(1)}(t,\beta_{0})}{\tilde{S}^{(0)}(t,\beta_{0})}-\frac{s^{(1)}(t,\beta_{0})}{s^{(0)}(t,\beta_{0})}\right\}dM_{i_{k}}(t,\beta_{0})= O_{P}(r^{-1})
	\end{equation*}
	\begin{proof}
		Note that
		\begin{equation}\label{eq: s2s0 minus s1s0}
			\begin{split}
				&\frac{1}{r}\sum_{k=1}^{r}\int_{0}^{\infty}\left\{\frac{\tilde{S}^{(1)}(t,\beta_{0})}{\tilde{S}^{(0)}(t,\beta_{0})}-\frac{s^{(1)}(t,\beta_{0})}{s^{(0)}(t,\beta_{0})}\right\}dM_{i_{k}}(t,\beta_{0})\\
				& = \frac{1}{r}\sum_{k=1}^{r}\int_{0}^{\infty}\left\{\frac{\tilde{S}^{(1)}(t,\beta_{0})-s^{(1)}(t,\beta_{0})}{s^{(0)}(t,\beta_{0})}\right\}dM_{i_{k}}(t,\beta_{0})\\
				&\quad+\frac{1}{r}\sum_{k=1}^{r}\int_{0}^{\infty}\left[\frac{s^{(1)}(t,\beta_{0})\{\tilde{S}^{(1)}(t,\beta_{0})-s^{(1)}(t,\beta_{0})\}}{s^{(0)}(t,\beta_{0})^2}\right]dM_{i_{k}}(t,\beta_{0})\\
				&\quad+\frac{1}{r}\sum_{k=1}^{r}\int_{0}^{\infty}\left[\frac{\{\tilde{S}^{(1)}(t,\beta_{0})-s^{(1)}(t,\beta_{0})\}\{\tilde{S}^{(0)}(t,\beta_{0})-s^{(0)}(t,\beta_{0})\}}{\tilde{S}^{(0)}(t,\beta_{0})s^{(0)}(t,\beta_{0})}\right]dM_{i_{k}}(t,\beta_{0})\\
				&\quad+\frac{1}{r}\sum_{k=1}^{r}\int_{0}^{\infty}\left[\frac{s^{(1)}(t,\beta_{0})\{\tilde{S}^{(0)}(t,\beta_{0})-s^{(0)}(t,\beta_{0})\}^{2}}{\tilde{S}^{(0)}(t,\beta_{0})s^{(0)}(t,\beta_{0})^{2}}\right]dM_{i_{k}}(t,\beta_{0}).
			\end{split}
		\end{equation}
		By Lemma \ref{lem: consist Sk}, the third and fourth terms on the right side of \eqref{eq: s2s0 minus s1s0} converge in probability to $0$ with convergence rate $r^{-1}$.
		For the first and second terms, by Lemma \ref{lem: consist Sk}, we have
		\begin{equation*}
			\begin{split}
				&\Big\|\frac{1}{r}\sum_{k=1}^{r}\int_{0}^{\infty}\left\{\frac{\tilde{S}^{(1)}(t,\beta_{0})-s^{(1)}(t,\beta_{0})}{s^{(0)}(t,\beta_{0})}\right\}dM_{i_{k}}(t,\beta_{0})\Big\|\\
				&\leq cr^{-1/2}\frac{1}{r}\sum_{k=1}^{r}\Big\|\int_{0}^{\infty}s^{(0)}(t,\beta_{0})^{-1}dM_{i_{k}}(t,\beta_{0})\Big\|
			\end{split}
		\end{equation*}
		and
		\begin{equation*}
			\begin{split}
				&\Big\|\frac{1}{r}\sum_{k=1}^{r}\int_{0}^{\infty}\left[\frac{\{\tilde{S}^{(1)}(t,\beta_{0})-s^{(1)}(t,\beta_{0})\}\{\tilde{S}^{(0)}(t,\beta_{0})-s^{(0)}(t,\beta_{0})\}}{\tilde{S}^{(0)}(t,\beta_{0})s^{(0)}(t,\beta_{0})}\right]dM_{i_{k}}(t,\beta_{0})\Big\|\\
				&\leq cr^{-1/2}\sup_{t\in[0,\tau]}\|s^{(1)}(t,\beta_{0})\|\Big\|\frac{1}{r}\sum_{k=1}^{r}\int_{0}^{\infty}s^{(0)}(t,\beta_{0})^{-1}dM_{i_{k}}(t,\beta_{0})\Big\|,
			\end{split}
		\end{equation*}
		where $c$ is a generic positive constant. In addition, we have $E\{\int_{0}^{\infty}s^{(0)}(t,\beta_{(0)})^{-1}dM(t,\beta_{0})\}=0$ by the martingale theory.
		Then we have $\Big\|r^{-1}\sum_{k=1}^{r}\int_{0}^{\infty}s^{(0)}(t,\beta_{0})^{-1}dM_{i_{k}}(t,\beta_{0})\Big\| = O_{P}(r^{-1/2})$ and hence the first and second terms on the right-side of \eqref{eq: s2s0 minus s1s0} converge in probability to $0$ with convergence rate $r^{-1}$.
		
	\end{proof}
\end{lemma}

\subsection*{Proof of Theorem 1}

Recalling the definition of $\tilde{\beta}_{\rm MCox}$, we have
\begin{equation}\label{eq: decomp}
	\begin{split}
		V_{h}^{-1/2}(\tilde{\beta}_{\rm MCox}-\beta_{0})
		&= V_{h}^{-1/2}\left(\tilde{\beta}_{\rm uni}-\beta_{0}-\Sigma_{0}^{-1}\Omega_{12}\Omega_{22}^{-1}\tilde{g}_{2}\right)\\
		&\quad- V_{h}^{-1/2}\left\{\widetilde{\Sigma}(\tilde{\beta}_{\rm uni})^{-1}\widetilde{\Omega}_{12}\widetilde{\Omega}_{22}^{-1}-\Sigma_{0}^{-1}\Omega_{12}\Omega_{22}^{-1}\right\}\tilde{g}_{2}.
	\end{split}
\end{equation}
We first prove the first term on the right side of \eqref{eq: decomp} converges to $N(0,I)$ in distribution.
By Taylor's expansion and some algebras, we have 
\begin{equation*}
	\begin{split}
		\tilde{\beta}_{\rm uni} - \beta_{0}
		&= \widetilde{\Sigma}(\bar{\beta})^{-1}\frac{1}{r}\sum_{k=1}^{r}\psi(Z_{i_{k}};\beta_{0})\\
		&\quad +\frac{1}{r}\sum_{k=1}^{r}\int_{0}^{\infty}\left\{\frac{\tilde{S}^{(1)}(t,\beta_{0})}{\tilde{S}^{(0)}(t,\beta_{0})}-\frac{s^{(1)}(t,\beta_{0})}{s^{(0)}(t,\beta_{0})}\right\}dM_{i_{k}}(t,\beta_{0}).
	\end{split}
\end{equation*}
where $\bar{\beta}$ is between $\beta_{0}$ and $\tilde{\beta}_{\rm uni}$.
Under Condition 1,  $\widetilde{\Sigma}(\bar{\beta})\stackrel{p}{\to}\Sigma_{0}$.		
Then by Lemma \ref{lem: remain rate} and Condition 3, we have
\begin{equation}\label{eq: beta al}
	V_{h}^{-1/2}(\tilde{\beta}_{\rm uni} - \beta_{0})= V_{h}^{-1/2}\Sigma_{0}^{-1}\frac{1}{r}\sum_{k=1}^{r}\psi(Z_{i_{k}};\beta_{0})+o_{P}(1).
\end{equation}
For $i=1,\dots,n$, let $\delta_{i}$ be the inclusion indicator for the $i$th sample, where $\delta_{i}=1$ if the $i$th sample is included and $\delta_{i}=0$ otherwise. Denote $\rho_{n} = r/n$ as the subsampling ratio.
Then we have
\begin{equation*}
	\begin{split}
		\tilde{g}_{2} 
		&= \frac{1}{n}\sum_{i=1}^{n}\left\{\rho_{n}^{-1}\delta_{i}-1\right\}\{h(Z_{i})-\mu_{0}\} +\left\{1-\rho_{n}^{-1}\frac{1}{n}\sum_{i=1}^{n}\delta_{i}\right\}\frac{1}{n}\sum_{i=1}^{n}\{h(Z_{i})-\mu_{0}\}.			
	\end{split}
\end{equation*}
By Chebyshev's inequality and Condition 2, we have $1-\rho_{n}^{-1}n^{-1}\sum_{i=1}^{n}\delta_{i}=O_{P}(r^{-1/2})$ and $n^{-1}\sum_{i=1}^{n}\{h(Z_{i})-\mu_{0}\}=O_{P}(n^{-1/2})$.
Then by Condition 3, we have
\begin{equation}\label{eq: gr}
	V_{h}^{-1/2}\tilde{g}_{2}=V_{h}^{-1/2}\frac{1}{n}\sum_{i=1}^{n}\left\{\rho_{n}^{-1}\delta_{i}-1\right\}\{h(Z_{i})-\mu_{0}\}+o_{P}(1).
\end{equation}
This together with \eqref{eq: decomp} and \eqref{eq: beta al} shows
\begin{equation*}
	V_{h}^{-1/2}(\tilde{\beta}_{\rm uni}-\beta_{0}-\Sigma_{0}^{-1}\Omega_{12}\Omega_{22}^{-1}\tilde{g}_{2})=\sum_{i=1}^{n}K_{i}+o_{P}(1),
\end{equation*}
where 
\begin{equation*}
	K_{i} = V_{h}^{-1/2}n^{-1}\Big[ \Sigma_{0}^{-1}\rho_{n}^{-1}\delta_{i}\psi(Z_{i};\beta_{0})-\Sigma_{0}^{-1}\Omega_{12}\Omega_{22}^{-1}\left\{\rho_{n}^{-1}\delta_{i}-1\right\}\{h(Z_{i})-\mu_{0}\}\Big].
\end{equation*} 
Now the problem reduces to prove $\sum_{i=1}^{n}K_{i}\to N(0,I)$ in distribution.
It is easy to verify that $\sum_{i=1}^{n}\var(K_{i})\to I$.
By Lindeberg-Feller central limit theorem in \cite{Vaart2000AS}, to prove $\sum_{i=1}^{n}K_{i}\to N(0,I)$ in distribution, it suffices to verify that for any given $\epsilon>0$, $\sum_{i=1}^{n}E\{\|K_{i}\|^2 1(\|K_{i}\|>\epsilon)\}\to 0$, 
where $1(\cdot)$ is an indicator function. Since for any given $\tau>0$, we have $E\{\|K_{1}\|^2 1(\|K_{1}\|>\epsilon)\}\leq E\{\|K_{1}\|^{2+\tau}/\epsilon^{\tau}\}$.
Then it suffices to verify that $nE(\|K_{1}\|^{2+\tau})=o(1)$ holds for some $\tau>0$.
Let $K_{1}^{(j)}$ be the $j$th element of $K_{1}$ for $j = 1,\dots,p$.
By the inequality $(t_{1}+t_{2})^{p}\leq 2^{p-1}(|t_{1}|^{p}+|t_{2}|^{p})$ and Condition 4, we have
\begin{equation*}
	\begin{split}
		nE(\|K_{1}\|^{2+\tau})
		& \leq c nE(|K_{1}^{(1)}|^{2+\tau}+\dots+|K_{1}^{(p)}|^{2+\tau})\\
		&\leq c n\left\{\var(K_{1}^{(1)})^{1+\tau/2}+\dots+\var(K_{1}^{(p)})^{1+\tau/2}\right\}.
	\end{split}
\end{equation*}
Since $E(K_{1}K_{1}^{\T})=n^{-1}I$, then we have $nE(\|K_{1}\|^{2+\tau})=o(1)$.
Under Condition 1, we have $\|\tilde{\beta}_{\rm uni}-\beta_{0}\|=O_{P}(r^{-1/2})$.
This together with Condition 2 and Slutsky's theorem proves $$\|\widetilde{\Sigma}(\tilde{\beta}_{\rm uni})^{-1}\widetilde{\Omega}_{12}\widetilde{\Omega}_{22}^{-1}-\Sigma_{0}^{-1}\Omega_{12}\Omega_{22}^{-1}\|=O_{P}(r^{-1/2}).$$ 		
Further by \eqref{eq: gr}, Chebyshev's inequality and Condition 2, we have $\tilde{g}_{2} = O_{P}(r^{-1/2})$.
Then recalling the definition of $V_{h}$ and by Condition 3, we have  $$V_{h}^{-1/2}(\widetilde{\Sigma}(\tilde{\beta}_{\rm uni})^{-1}\widetilde{\Omega}_{12}\widetilde{\Omega}_{22}^{-1}-\Sigma_{0}^{-1}\Omega_{12}\Omega_{22}^{-1})\tilde{g}_{2} = o_{P}(1).$$

\subsection*{Proof of Theorem 2}

Recalling the definition of $V_{h}$, we have
\begin{equation*}
	\begin{split}
		V_{h}
		&= r^{-1}\Sigma_{0}^{-1}(\Sigma_{0}-\Omega_{12}\Omega_{22}^{-1}\Omega_{21})\Sigma_{0}^{-1}\\
		& = \frac{1}{n}\Sigma_{0}^{-1}E\left(\left[\rho_{n}^{-1}\delta \psi(Z;\beta_{0})-\Omega_{12}\Omega_{22}^{-1}\left\{\rho_{n}^{-1}\delta-1\right\}\{h(Z)-\mu_{0}\}\right]^{\otimes2}\right)\Sigma_{0}^{-1}.
	\end{split}
\end{equation*}
Without loss generality,  we assume $\mu_{0}=0$ since $V_{h}$ is invariant if $h$ is replaced by $h-c$ for any $q$ dimensional constant vector $c$.
Then we have
\begin{align*}	
	&E\left(\left[\rho_{n}^{-1}\delta \psi(Z;\beta_{0})-\Omega_{12}\Omega_{22}^{-1}\left\{\rho_{n}^{-1}\delta-1\right\}h(Z)\right]^{\otimes2}\right)\\
	&=E\left\{\psi(Z;\beta_{0})^{\otimes2}\right\}+E\left[\left\{\rho_{n}^{-1}-1\right\}\{\psi(Z;\beta_{0})-\Omega_{12}\Omega_{22}^{-1}h(Z)\}^{\otimes2}\right]\\
	& \geq E\left\{\psi(Z;\beta_{0})^{\otimes2}\right\} = \Sigma_{0}.
\end{align*}
Then $V_{h}$ attains the minimum if and only if $\psi(z;\beta_{0})=\Omega_{12}\Omega_{22}^{-1}h(z)$. 
Recalling the definition of $\Omega_{12}$ and $\Omega_{22}$, it is easy to verify that $\psi(z;\beta_{0})=\Omega_{12}\Omega_{22}^{-1}h(z)$ if and only if  $\psi(z;\beta_{0})=M h(z)$ for some matrix $M$.

\section{Marginal Risk of Maternal Age}

\begin{figure}
	\centering
	\includegraphics[scale=0.4]{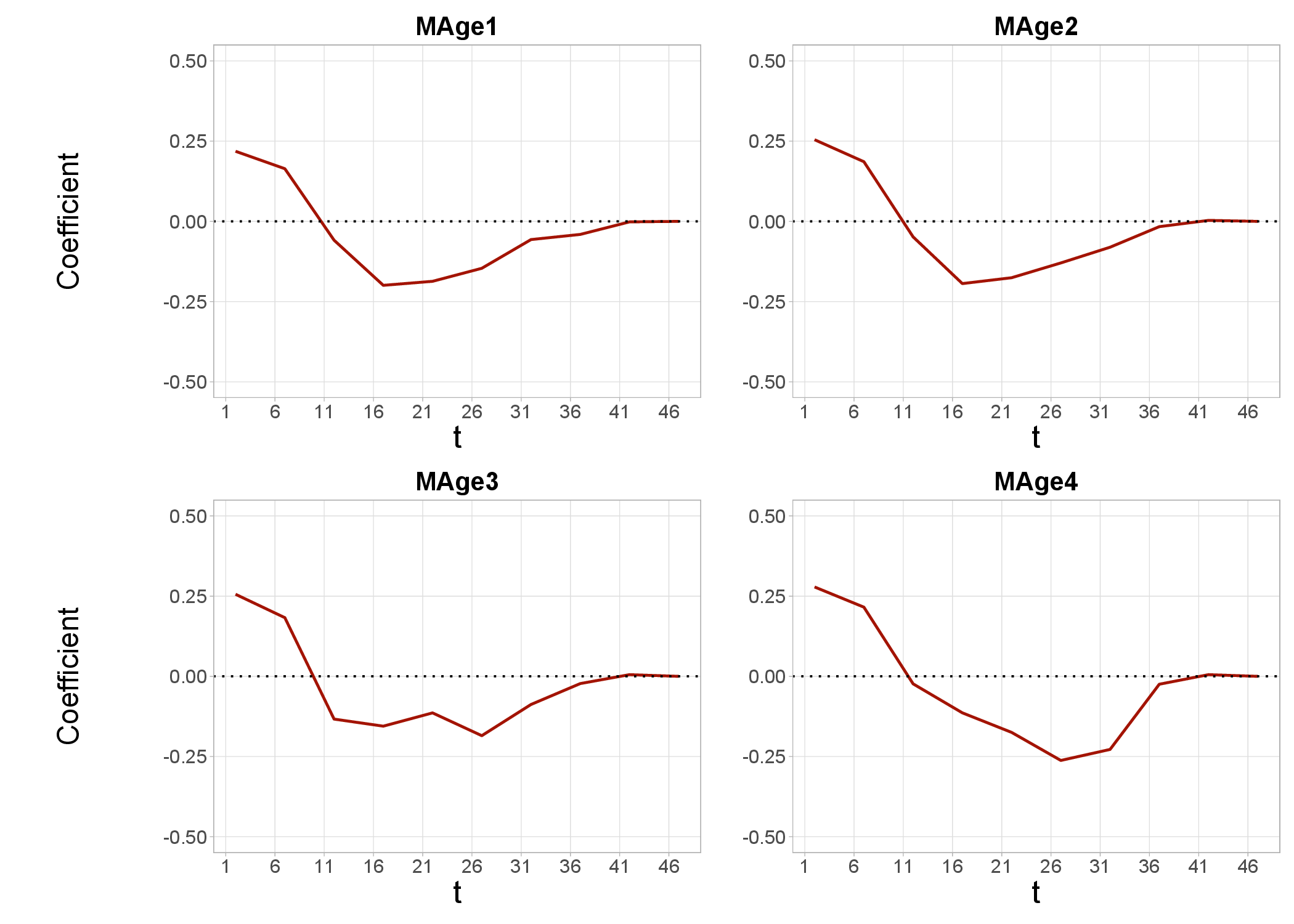}
	\caption{Marginal risk of maternal age on fetal death across different gestation stage. MAge: maternal age (MAge1: 20-24 years, MAge2: 25-29 years, MAge3: 30-34 years, MAge4: 35 years and older)}
	\label{fig: coef marginal}
\end{figure}

We use a pilot sample of size 10000 to plot the marginal risk of maternal age across different gestational stages. As shown in Figure \ref{fig: coef marginal}, the marginal risk of maternal age on fetal death exhibits an approximately quadratic variation with gestational weeks.
This observation motivates the adoption of a quadratically time-varying coefficient model in our analysis.

\bibliographystyle{asa}
\bibliography{MCox-arXiv-Ref}
\end{document}